\documentclass{emulateapj}
\received{}
\accepted{}
\slugcomment {Submitted to ApJL; preprint NSF-KITP-07-138}
\begin{document}
\title{The Progenitors of Type Ia Supernovae}
\author{Christopher J. Pritchet\altaffilmark{1}, D. Andrew Howell\altaffilmark{2}, and Mark Sullivan\altaffilmark{3,2}}


\altaffiltext{1}{Department of Physics and Astronomy, University of Victoria, PO Box 3055, Stn CSC, Victoria, BC V8W 3P6, Canada; pritchet@uvic.ca}
\altaffiltext{2}{Department of Astronomy, University of Toronto, 50 St. George St., Toronto, ON M5S 3H4, Canada; howell@astro.utoronto.ca}
\altaffiltext{3}{Department of Physics (Astrophysics), University of Oxford, Denys Wilkinson Building, Keble Road, Oxford, OX1 3RH, United Kingdom; sullivan@astro.ox.ac.uk}

\begin{abstract}

Type Ia supernovae (SNe Ia) occur in both old, passive galaxies and
active, star-forming galaxies. This fact, coupled with the strong
dependence of SN Ia rate on star formation rate, suggests that SNe Ia
form from stars with a wide range of ages.  Here we show that the
rate of SN Ia explosions is about 1\% of the stellar death rate,
independent of star formation history. The dependence of SN Ia rate on
star formation rate implies a delay time distribution proportional to
$t^{-0.5\pm0.2}$. The single degenerate channel for SNe Ia can be made
to match the observed SN Ia rate -- SFR relation, but only if white
dwarfs are converted to SNe Ia with uniform efficiency of $\sim$1\%,
independent of mass. Since low-mass progenitors are expected to have
lower conversion efficiencies than high mass progenitors, we
conclude that some other progenitor scenario must be invoked to
explain some, or perhaps all, SNe Ia.

\end{abstract}

\section{Introduction}

Type Ia supernovae (SNe Ia) are, next to gamma ray bursts, among the
most powerful transient objects in the Universe. SNe Ia are widely
believed to be explosions of C+O white dwarfs (WDs) whose mass has
grown by accretion to the Chandrasekhar mass of 1.4 M$_\odot$
\citep{chandra1931}. Evidence for this includes total energy
released (consistent with conversion of C+O to Fe), lack of hydrogen
in the maximum light spectra (consistent with an object whose envelope
has been lost), light curve shape (broadly in agreement with energy
deposition from radioactive decay of Fe-peak elements, as expected for
exploding C+O WD’s), and the occurrence of SNe Ia in old ($\sim$10
Gyr) stellar populations containing only low mass stars (for which
exploding WDs represent the natural, and perhaps only possible, SN Ia
progenitor).

Beyond this consensus on the exploding WD model, the precise
nature of the progenitors of SNe Ia remains poorly
constrained. Progenitor models are broadly classified as
``single-degenerate" (SD), in which a white dwarf grows in mass by
accretion from an evolving binary companion \citep{whel1973}, and
``double degenerate" (DD), in which two white dwarfs merge
\citep{iben1984,webb1984}. A variety of evolutionary paths may lead to SD and
DD events \citep[e.g.][]{livio2001,yung2005,partha2007,greggio2005}.

The fact that SNe Ia are observed in elliptical galaxies has led to
the persistent and widespread presumption that SN Ia progenitors are
old, low-mass stars. Yet over the years there have been several hints
suggesting that the SN Ia rate is enhanced in star-forming galaxies
\citep[e.g.][]{oemtins1979,vdb1990,dellaval1994}.  Two recent papers
\citep{mann2005,sull2006} have marshalled evidence for a strong
dependence of the SN Ia rate on star formation rate (SFR), thus
linking SNe Ia with young, massive stellar populations.
\citet{mann2005} made use of a sample of low redshift supernovae
discovered in 5 targeted searches \citep{capp1999}, and demonstrated a
strong dependence of specific SN Ia rate (rate per unit mass) on
morphology or galaxy colour. This result was confirmed and extended by
\citet{sull2006}, who analyzed a sample of spectroscopically-confirmed
SNe Ia (redshift z=0.2-0.75, spanning cosmic time 7-11 Gyr after the
Big Bang) from the Supernova Legacy Survey \citep{astier2006} (SNLS).

The SN Ia rate -- SFR correlation found by \citet{sull2006} 
is shown in the lower panels of Fig. 1 (both axes are normalized
by galaxy mass). Passive galaxies are plotted as a single data point
in the lower left hand panel; most of these galaxies are old. The variation
of SN Ia rate from passive galaxies to the most active galaxies is
dramatic: passive galaxies possess a mass-normalized SN Ia rate that
is a factor of ten or more lower than found for active star-bursting
systems. 

In this paper we present a simple model for the SN Ia rate -- SFR
relation in terms of stellar evolutionary timescales. This model,
which is based on the SD scenario, allows an estimation of the
efficiency of conversion of white dwarfs into SNe Ia. This provides,
as we shall see, a constraint on the nature of SN Ia progenitors.

\section {The Model}

\subsection {White Dwarf Formation Rates}

A first step towards understanding the rate
of SNe Ia in galaxies is a calculation of the rate at which WDs are
formed in different stellar populations. It is well-known that WDs (M
$<$ 1.4 M$_\odot$) form as the end-point of stellar evolution for
stars with initial masses M $<$ 8 M$_\odot$ \citep[e.g.][and
references therein]{weid2000,kal2007}. The rate at which white dwarfs
form clearly depends on evolutionary time scales as a function of
initial stellar mass, and the initial mass function or IMF. For a
\citet{salp1955} mass function $dN/dm \propto m^a$ ($a=-2.35$), and a
power-law approximation to evolutionary timescales $t_{evol} \simeq
10^{10} (M/M_\odot)^b$ yr ($b=-2.5$), it can be shown that white
dwarfs form from a pure instantaneous burst of star formation at a
rate that decreases with time roughly as $t^{(a-b+1)/b} \sim t^{-1/2}$. 
This (somewhat counterintuitive) result is due to the fact
that the evolutionary timescales of massive stars are so much shorter
than for low mass stars, even in the presence of a steeply-sloped IMF
favouring low mass stars.

\begin{figure}
\begin{center}
\includegraphics[angle=-90,scale=0.30]{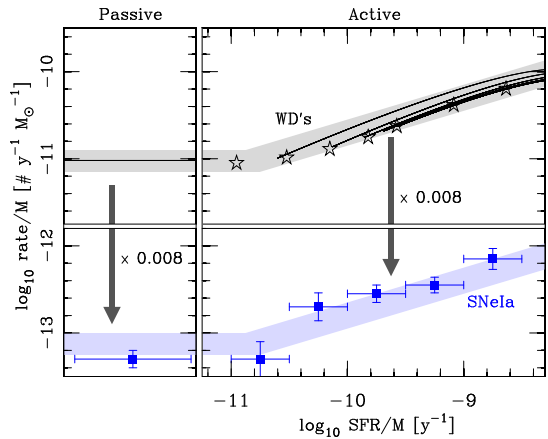}
\end{center}
\caption {Rate of formation of white dwarfs and Type Ia supernovae, as
a function of star formation rate; all rates are normalized
per unit luminous stellar mass. The left panels are for passive
(non-star-forming) galaxies (E/S0's) with an age of 8 Gyr (solid line)
and extending from 5 to 11 Gyr (shaded area). The right panels are for
objects with star formation.
At the bottom of the diagram, data points
show the observed formation rate per unit host galaxy mass of Type Ia
supernovae from the SNLS survey \citep{sull2006}, as a function of star formation
rate per unit mass. 
The upper shaded area gives the locus of
models (with a wide range of ages and star formation rate histories)
for the formation rate of white dwarfs. The solid lines are models
with SFR $\propto t^{-p}$ , with $p=0.95$ (rapidly decreasing SFR) on top and $p=-1$
(rapidly increasing SFR) at the bottom. Age runs along each model
line, from 0.1 Gyr at upper right to 10 Gyr at lower left. The star
symbols are composite stellar populations (old bulge/halo plus
star-forming disk, with proportions chosen to match observed colours
along the Hubble sequence \citep{buzz2005}; from left to right the symbols
represent Sa, Sb, Sc, Sd, and 3 starburst galaxies ($p=-0.8$) with ages
10 Gyr, 3 Gyr, and 1 Gyr. Scaling the formation rate of white dwarfs by a
factor $\eta_{obs}=0.008$ provides an excellent match to the observations at
all star formation rates (including passive galaxies). }
\end{figure}

The results of a detailed calculation of WD
formation rate are shown in the upper panels of Fig. 1, using
more exact evolutionary timescales \citep{buzz2002}, and a power-law
SFR $\propto  t^{-p}$ \citep[e.g.][]{buzz2005}. (Other SFR(t) functions give equivalent
results, as will be shown.) For passive
systems with no star formation (left hand
panels), we plot the WD formation rate per unit galaxy mass
(WDR/M) for a star burst with age 7.9 Gyr (the approximate age of the
Universe at the mean redshift of the SNLS SNe). The shaded region in
Fig. 1 includes systems as young as 5 Gyr (which would have formed $\geq2$
Gyr after the big bang for the most distant objects in SNLS).  For
active systems with star formation, we plot WDR/M as a function of
star formation rate per unit galaxy mass (SFR/M).
 
Remarkably, the form of this diagram is very insensitive to the
details of star formation history (other than age).  Old starburst
models with increasing SFR(t) ($p<0$) overlap young models with
decreasing SFR ($p>0$), and all of these models lie within the upper
shaded region of the figure. A variety of composite stellar
populations, comprising an old passive bulge/halo and a star-forming
disk \citep[with the proportion of bulge to disk mass given by][to
match observed galaxy colours]{buzz2005} can be seen to fall in the shaded
region of Fig. 1, as do models with bursts of star formation. Models
with exponentially decreasing SFR(t) match the locus of the SFR $\propto t^{-p}$
models, as do models constructed using the PEGASE.2 evolutionary infall
scenario \citep{leborgne2002}. Finally we note that the range of B-K colours exhibited
by all of these models spans the full dynamic range of colour
exhibited by real galaxies. We conclude that the locus of the white
dwarf formation rate shown in Fig. 1 is well-determined, independently
of uncertainties in the evolutionary models.
 
SFR/M has units of $T^{-1}$, and is approximately the inverse of the
gas consumption timescale of a galaxy (if SFR were held
constant). Identifying $\tau \simeq$ (SFR/M)$^{-1}$ as a
characteristic mean age of a model, one expects (WDR/M) $\simeq
\tau^{-1/2} \simeq $ (SFR/M)$^{1/2}$, roughly as observed in
Fig. 1. The increase of WD formation rate with SFR/M is therefore a
reflection of the fact that the mean age of stellar populations with a
large SFR/M (small gas consumption timescale) is smaller.

\subsection {Type Ia Supernova Rates and the Conversion Efficiency}

These calculations of specific white dwarf formation rate, WDR/M, can
now be compared with estimates of the SNe Ia rate per unit stellar
mass as derived from the SNLS survey \citep{sull2006}. Scaling the white dwarf
formation rate curve by a factor $\eta_{obs}=0.008$ provides an excellent
match to the SNLS observations \citep[and also to the observations of][]{mann2005}. 

However, to convert $\eta_{obs}$ (the ratio of SN Ia to WD
formation rates) to the  physically more interesting conversion
efficiency $\eta$ (the fraction\footnote{Note that $\eta$
includes the binary fraction -- e.g. if the fraction of binaries were
0.5, then the fraction of WD’s in binaries that become SNe Ia would be
$\sim$0.02. Available evidence \citep{lada2006} suggests that the fraction of binaries
increases with mass.} of white dwarfs that explode as SNe
Ia), it is necessary to correct for the time delay, $\Delta
t_{WD-SN}$, between the formation of a WD primary and an SN Ia
event. In the case of an SD system, this delay is due to the residual time that
it takes the secondary star (of mass $M_2= q M_1$, where $q$ is mass
ratio) to evolve. The available evidence suggests that the
distribution of $q$ is flat \citep{lars2001}, or even strongly peaked
towards equal masses \citep{pins2006}, for close binaries. 

We have computed detailed models for SD binaries with different
distributions of q, using both numerical integration and Monte Carlo
techniques. A flat distribution in $q$ results in rates that are
$<2\times$ lower than for “undelayed” $q=1$ models (see
Fig. 2). Most important, this factor is nearly constant for all SFR/M.
Correcting $\eta_{obs}$ for the distribution of $q$ values yields a
white dwarf to SN Ia conversion efficiency $\eta$ in the range
0.01-0.015.  (A power-law distribution of mass ratios $N(q) \propto q^x$,
corresponding to a distribution of secondary masses
$N(M_2) \propto M_2^x$, results in constant $\eta$
provided that $x > -1.5$. A distribution of secondary masses drawn from the
IMF gives WDR/M$\simeq$constant, independent of SFR/M. 
This is discussed further in \S 3 and in
\citet[][Paper 2]{pritchet2008}.)

A calculation for the DD model is beyond the scope of this
paper. However we note that a similar result ($\eta=$constant) is
obtained if the distribution of semimajor axes for DD binaries is
independent of mass (Paper 2). We also note that $\eta \approx 0.01$
is valid for any mass range dominated by the SD channel.

An alternate way of describing the data is through the delay time
distribution (DTD). From \S 2, it can be seen that the data are
consistent with a continuous delay time distribution DTD $\propto
t^{-0.5}$ (for a burst). Fig. 3 shows that the data are bracketed by
DTD $\propto t^{-0.3}$ and $t^{-0.7}$.

\begin{figure}
\begin{center}
\includegraphics[angle=-90,scale=0.30]{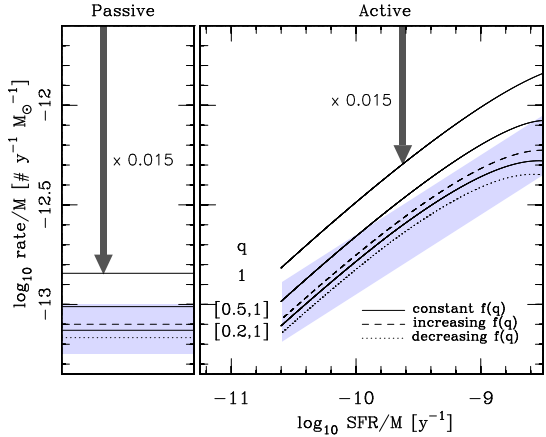}
\end{center}
\caption {
Effect of a distribution of secondary masses for the SD model. The
shaded region is from Fig. 1. The models all have $\mu_{obs}=0.015$
(double that in Fig. 1) and SFR $\propto t^{-0.95}$. The top solid
line is for binaries with mass ratio $q=1$ (as in Fig. 1 -- no delay
between WD and SN). The other two solid lines are for populations of
binaries with a flat distribution of $q$ over the range [0.5,1]
(middle), and [0.2,1] (bottom). The dashed/dotted lines are mildly
increasing/ decreasing (by a factor of 2) $q$ distributions over
$q=$[0.2,1]. It can be seen that the effect of a distribution of mass
ratios is to decrease SN rates at a given SFR by at most a factor of
$\sim$2. } 
\end{figure}

In summary: The SNIa rate is approximately 1\% of the stellar
death rate, from passive galaxies (SFR=0) to starburst
galaxies (SFR/M $\geq 10^{-9}$ y$^{-1}$), with little dependence on
the details of the stellar population mix.  The dependence of SN Ia
rate on star formation rate in both active and passive galaxies is
consistent with a simple timescale model based on the SD scenario,
with a uniform WD conversion efficiency $\eta\approx$1\% for all
stars. The SN Ia rate -- SFR rate dependence is also consistent with a
continuous DTD $\propto t^{-0.5\pm0.2}$.  If more than one channel
produces SNe Ia, $\eta\approx1$\% still holds for any progenitor
mass range in which the SD channel is dominant.

Our models are different from those of \citet{greggio2005} since we
consider the scaling of the SN Ia rate with SFR, thus minimizing
the effects of variations in star formation history. In addition our
models do not initially include variations of $\eta$ with mass (for
reasons that will become apparent). However, the slope of the DTD
for a simple aging burst agrees with that of \citet[][Fig. 3]{greggio2005}
for $t < 1$ Gyr.
 
\subsection {Uncertainties in the Conversion Efficiency}

What is the uncertainty in the efficiency factor $\eta$? Changing the
functional form of SFR(t) has already been eliminated as a source of
error. Changing the IMF to a two-piece or three-piece IMF
\citep{kroupa2007} (with increasing power-law slope at higher mass)
remains consistent with the locus of white dwarf models in Fig. 1,
provided $\eta$ is reduced by 25\%. Systematic errors in mass affect
the vertical placement of data points in Fig. 1, and these could
be a factor of 2 or larger.  Systematic errors of a
factor of 2 in SFR/M affect $\eta$ by about 40\%. The overall
completeness of the SNLS rates, from which $\eta$ was derived,
approaches 100\% \citep{sull2006}. It thus appears that the efficiency
factor $\eta$ is accurate to about  a factor of two. Most
important, the relative efficiency factor between galaxies of
different SFR/M should be unaffected by these systematic errors.
 
\subsection {Comparison with Other Values of Conversion Efficiencies}

Is $\eta=0.01$ reasonable from other considerations? This is a
difficult question to answer. Theoretical models
\citep[e.g.][]{yung2005} give efficiencies in the range 0.001--0.01,
but are subject to a host of uncertainties (e.g. the distributions of
binary mass ratios and separations, and the precise evolutionary
scenario for SNe Ia). 
For the Milky Way, the observed SN Ia rate \citep{capp2001} and
WD formation rate \citep[cf.][]{phill2002,Soker2006} yield
$\eta \simeq$ 0.1--2\%. This number is extremely uncertain, but it
would seem to rule out conversion efficiencies larger than a few per
cent.

Other observational determinations of efficiency (from the ratio of
SNe Ia to core collapse SNe, from the A and B rate parameters
\citep{scan2005}, and from abundances) span a huge range (1--40\%),
even approaching 100\% for intermediate mass stars
\citep{maoz2007}. Part of this discrepancy is due to the
different progenitor mass range (3--8M$_\odot$) used by Maoz; however
our determination of $\eta=0.01$ will still apply for any mass range
dominated by the SD channel. To raise the value of $\eta$ to 0.1
would require 90\% SN Ia incompleteness in the SNLS survey, or SFR
values that are overestimated by a factor of 100. Both of these
possibilities are untenable.

\section {Implications for SN Ia Progenitors}

We now turn to
relative values of $\eta$ for populations with different SFR. The relative
WD to SN Ia conversion efficiency $\eta$ can be measured directly in Fig. 1
for populations with different SFR/M, i.e. mean age or mass. In
particular, it can be seen that the conversion efficiency for old
stars in passive galaxies is similar to that for the most active
starburst galaxies. The variation of $\eta$ with SFR is smaller than a
factor two.
 
The SD model results in lower conversion
efficiencies $\eta$ for low mass stars, for a variety of reasons. A
passive 10 Gyr old population is populated with stars with M $\leq$ 1
M$_\odot$, and, because of mass loss on the giant branch, such stars
produce white dwarfs of mass 0.5 M$_\odot$ \citep{weid2000,kal2007}. A binary
system initially comprising two 1 M$_\odot$ stars would therefore not
evolve into a 1.4 M$_\odot$ white dwarf unless mass transfer were
exceptionally efficient. Below about 2 M$_\odot$ (turnoff age 1 Gyr),
a He (rather than C+O) white dwarf is the likely outcome in a close
binary \citep{greggio2005}, and such stars do not produce SNe Ia. 
The evolutionary
time of the secondary sets the clock for the SN Ia explosion, and at
low mass, some fraction of the secondaries have not yet completed
their evolution. An increasing binary fraction with mass
\citep{lada2006} lowers the SN Ia rate at low mass relative to high
mass. All of these factors will conspire to produce lower efficiencies
at low SFR/M, relative to the most active galaxies.
 
The models of \citet{greggio2005} represent SD efficiency at low mass
by including a continuous variation of $\eta$ for $M_2 <$ 2 M$_\odot$,
with $\eta_{1M_\odot}/\eta_{2M_\odot} < 0.1$. The result of such an
efficiency variation is shown in Fig. 3. Normalizing the models to the
most active galaxies, the computed and observed SN Ia rates for
passive galaxies disagree by a factor of 7. If SD progenitor scenarios
are less efficient at low mass, as is expected, then SD progenitors
cannot be the only objects producing SNe Ia; some other class of model
(perhaps the double degenerate model) must be invoked to explain some, and perhaps
all, SNe Ia. A similar conclusion has been reached by, for example,
\citet{greggio2005} and \citet{yl2000}. This conclusion is also consistent
with the observational lack of hydrogen in the late-time ejecta of SNe Ia
\citep{leon2007}.

If secondary star masses for {\it close}
binaries were drawn from the IMF, then close binaries with a high mass primary
and low mass secondary would be common, and the number of SNe Ia in
old progenitors would be enhanced, compensating for the 
expected lower conversion efficiency at $M < 2 M_\odot$.  
We reiterate, however, that the available evidence supports an IMF
for secondaries in close binaries that is {\it not} drawn from the
IMF \citep[e.g.][]{lars2001,pins2006}.

\begin{figure}
\begin{center}
\includegraphics[angle=-90,scale=0.30]{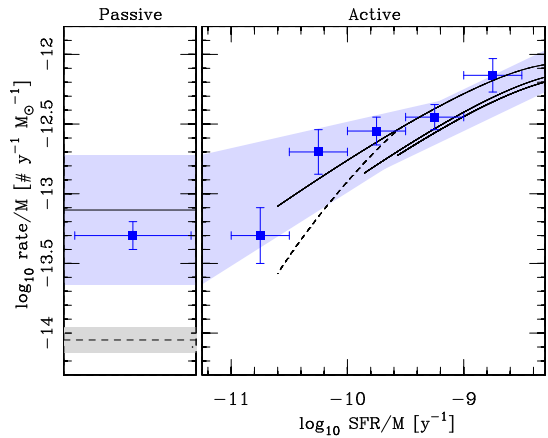}
\end{center}
\caption { Data from \citet{sull2006} is
compared with SFR $\propto t^{-p}$ models. 
Solid lines are models from Fig. 1.
These models correspond to a delay time distribution DTD roughly $\propto t^{-0.5}$.
The upper shaded area is bounded by DTD$\propto t^{-0.3}$ (top boundary) 
and $t^{-0.7}$ (bottom boundary); these models are normalized at high SFR.
The dashed line represents a $p=0.95$ model
with the WD-SN efficiency at 1 M$_\odot$ reduced by a factor of 10. The
dashed horizontal line and shaded area at the lower left shows the
predictions of the low efficiency model for the passive galaxies. It
can be seen that models that are normalized correctly for starburst
galaxies fail by a factor of 7 or more to account for the SN Ia rate
in passive galaxies.}
\end{figure}

The most sub-luminous SNe Ia (stretch parameter $s<0.8$) are neglected
in the rates \citep{sull2006}. Including these objects (mostly in
passive galaxies) makes the disagreement between observations and
predictions of the SD scenario even more extreme. A strong
type-dependent incompleteness (with 90\% incompleteness for starburst
galaxies) would be required to explain the observations; there is no
evidence for such an extreme effect.

\section {Conclusions}

The SN Ia rate is about 1\% of the white dwarf formation rate,
for all star formation rates and for a wide range of stellar population
mixes.  The SN Ia rate is also consistent with a continuous delay time
distribution $\sim t^{-0.5\pm0.2}$. The dependence of the SN Ia rate
on SFR matches the predictions of a simple timescale model based on
the single degenerate channel, but only under the (unrealistic)
assumption that the fraction (1\%) of white dwarfs exploding as SNe Ia
is independent of progenitor mass. We conclude, as have others
\citep[e.g.][] {greggio2005,yl2000}, that a class of model other than
the SD model must be invoked to explain at least some SNe Ia. Nevertheless,
the conversion efficiency of 1\% holds for any mass range dominated by the SD channel.
This conclusion is valid provided that secondary stars in close binaries
have a distribution of mass ratios $q$ that is approximately flat or rising towards
$q=1$. Clearly there is a need for a better understanding of the the
distribution of secondary masses in close binaries.

Finally, it should be noted that the predictions of SN Ia rate in
this paper provide a better match to observations than does the empirical
$A \cdot M + B \cdot SFR$ formula \citep{scan2005,sull2006}.
The implications of this will be discussed
in Paper 2.

\acknowledgments
We are grateful to the hospitality of the Kavli Institute of
Theoretical Physics (UCSB), and to Lars Bildsten, Rosanne Di Stefano,
Philipp Podsiadlowski, Evan Scannapieco, Lev Yungel’son, Sidney van
den Bergh, David Hartwick, Don VandenBerg, and David Branch for
helpful discussions. We also acknowledge the insightful comments of an anonymous
referee.  This work
was supported by the Natural Sciences and Engineering Research Council
of Canada, and in part by the National Science Foundation under grant
no. PHY05-51164. MS acknowledges support from the Royal Society.

\end{document}